\documentclass{PoS}

\title{Dark Stars: Begynnelsen}

\ShortTitle{Dark Stars: Begynnelsen}

\author{\speaker{Paolo Gondolo}$^1$,
Katherine Freese$^2$, 
Douglas Spolyar$^3$, 
Anthony Aguirre$^3$, 
Peter Bodenheimer$^3$, 
Jeremy A. Sellwood$^4$, and
Naoki Yoshida$^5$\\
        $^1$University of Utah, $^2$University of Michigan, $^3$University of California, Santa Cruz, $^4$Rutgers University, $^5$University of Tokyo\\
        E-mail: \email{paolo@physics.utah.edu}, \email{ktfreese@umich.edu}, \email{dspolyar@physics.ucsc.edu}, \email{aguirre@scipp.ucsc.edu}, \email{peter@ucolick.org}, \email{sellwood@physics.rutgers.edu}, \email{nyoshida@a.phys.nagoya-u-ac.jp} }

\abstract{The first phase of stellar evolution in the history of the universe
may be Dark Stars, powered by dark matter heating rather than by fusion.
Weakly interacting massive particles, which are their own
antiparticles, can annihilate and provide an important heat source for
the first stars in the universe.  This and the following contribution present the story of Dark Stars. In this first part, we describe the conditions under which dark stars form in the early universe: 
1) high dark matter densities, 2) the annihilation products get
stuck inside the star, and 3) dark matter heating wins over all other cooling
or heating mechanisms. 
}

\FullConference{Identification of dark matter 2008\\
		 August 18-22, 2008\\
		 Stockholm, Sweden}

\begin{document}

We have studied the effect
of Dark Matter particles on the very first stars to form in the universe.
We have found a new phase of stellar evolution: the first stars to form
in the universe may be ``Dark Stars,''
powered by dark matter annihilation rather than nuclear fusion.
We first reported on this work in
(\cite{SpolyarFreeseGondolo08}).

The Dark Matter particles we considered are Weakly Interacting
Massive Particles (WIMPs) (such as the Lightest Supersymmetric Particle),
which are one of the major motivations for building the Large
Hadron Collider at CERN that will begin taking data very soon.
These particles are their own antiparticles; they annihilate among
themselves in the early universe, leaving the correct relic 
density today to explain the dark matter in the universe.  These particles will 
similarly annihilate wherever the DM density is high.
The first stars are particularly good sites for annihilation
because they form at high redshifts (density scales as $(1+z)^3$) and
in the high density centers of DM haloes.  
The first stars form at redshifts $z \sim 10-50$ 
in dark matter (DM) haloes
of $10^6 M_\odot$ (for reviews see e.g. 
\cite{RipamontiAbel05,BarkanaLoeb01,BrommLarson03}; see also
\cite{Yoshida_etal06}.)
One star is thought to form inside one such
DM halo. 

As canonical values for the particle peoperties, we will use the standard annihilation 
cross section,
$\langle \sigma v \rangle = 3 \times 10^{-26} {\rm cm^3/sec}$,
and a particle mass $m_\chi = 100$ GeV;
but we will also consider a broader range of 
masses and cross-sections.
In (\cite{SpolyarFreeseGondolo08}) we found that
DM annihilation provides a powerful heat source in the first stars, a
source so intense that its heating overwhelms all cooling mechanisms;
subsequent work has found that the heating dominates over fusion as well
once it becomes important at later stages (see accompanying contribution \cite{DarkStarsII}). 
Paper I (\cite{SpolyarFreeseGondolo08})
suggested that the
very first stellar objects might be ``Dark Stars,'' a new phase of
stellar evolution in which the DM -- while only a negligible fraction
of the star's mass -- provides the power source for the star through
DM annihilation.

\begin{figure}[b]
\centerline{\includegraphics[width=0.5\textwidth]{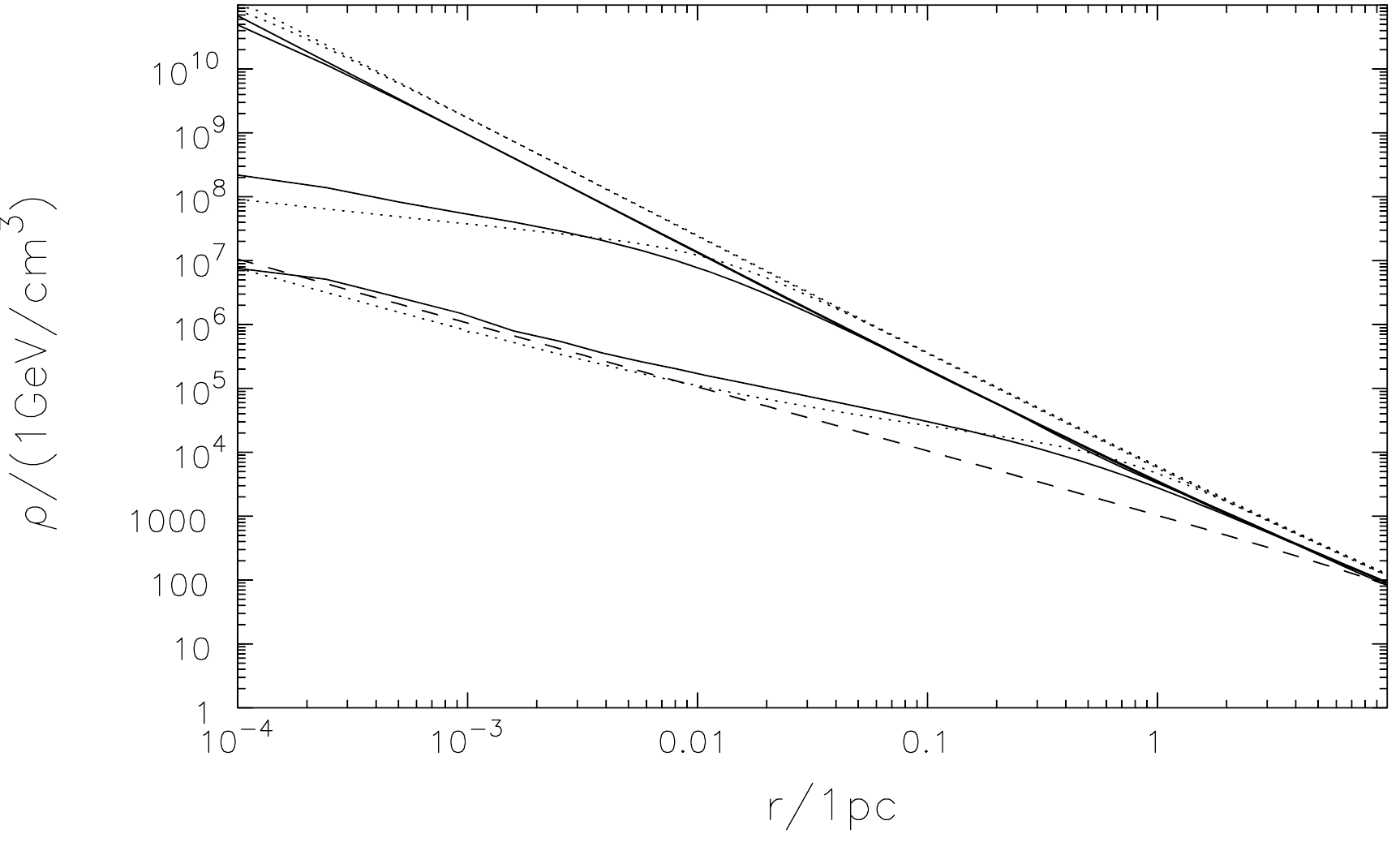}}
\caption{Adiabatically contracted DM profiles in the first
protostars for an initial NFW
profile (dashed line) using (a) the Blumenthal method (dotted lines)
and (b) an exact calculation using
Young's method (solid lines), for $M_{\rm vir}=5 \times 10^7
M_\odot$, $c=2$, and $z=19$.  The four sets of curves 
correspond to a baryonic core density of $10^4, 10^8, 10^{13},$ and
$10^{16}{\rm cm}^{-3}$.  The two different
approaches to obtaining the DM densities find values that differ by
less than a factor of two.
}
\end{figure}

\section{Three Criteria}

Paper I (\cite{SpolyarFreeseGondolo08})
outlined the three key ingredients for Dark Stars:
1) high dark matter densities, 2) the annihilation products get
stuck inside the star, and 3) DM heating wins over other cooling
or heating mechanisms.  These same ingredients are required throughout
the evolution of the dark stars, whether during the protostellar
phase or during the main sequence phase.

{\bf First criterion: High dark matter density inside the star.}
To find the DM density profile, we start with an 
overdense region of $\sim 10^6 M_\odot$
with an NFW (\cite{NavarroFrenkWhite96}) 
profile for both DM and gas, where the gas contribution is
15\% of that of the DM. Originally we used adiabatic contraction ($M(r)r$ =
constant) (\cite{Blumenthal_etal85}) and matched
onto the baryon density profiles given by 
\cite{AbelBryanNorman02}
and \cite{Gao_etal07} to
obtain DM profiles.  This method is overly simplified:
it considers only circular orbits of the DM particles. 
Our original DM profile matched
that obtained numerically in
\cite{AbelBryanNorman02}
with $\rho_\chi \propto r^{-1.9}$, for both their 
earliest and latest profiles; see also 
\cite{NatarajanTanO'Shea08} for a recent
discussion.  Subsequent to our original work, 
we have done an exact calculation (which includes
radial orbits) 
(\cite{FreeseGondoloSellwoodSpolyar08})
and found that our original results were remarkably 
accurate, to within a factor of two.  Our resultant 
DM profiles are shown in Fig.~1.
At later stages, we also consider possible further enhancements
due to capture of DM into the star (see \cite{DarkStarsII}).

\begin{figure}[t]
\centerline{\includegraphics[width=0.5\textwidth]{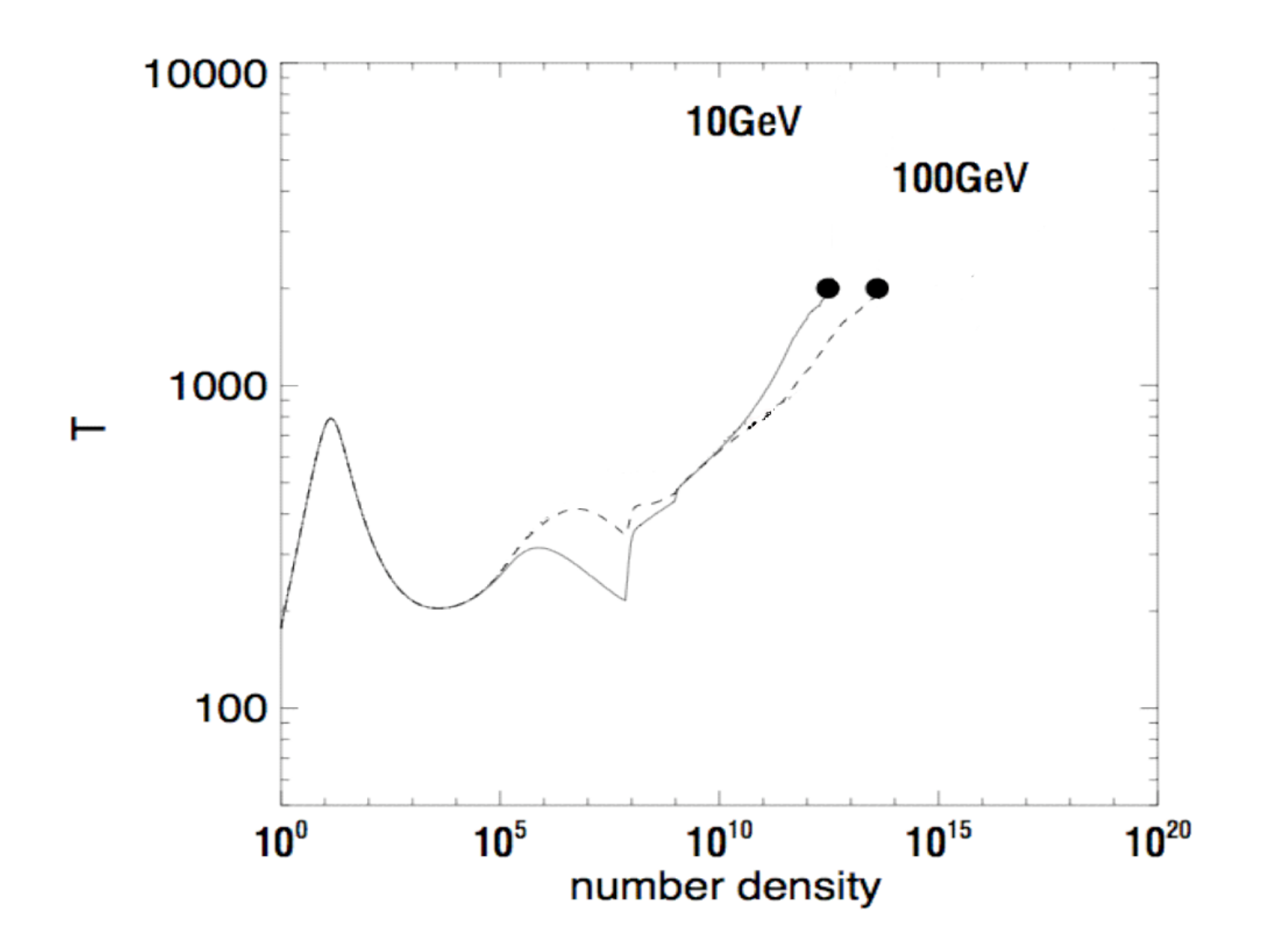}}
\caption{ Temperature (in degrees K) as a function of hydrogen density 
(in cm$^{-3}$) for the first protostars, with DM annihilation included,
for two different DM particle masses (10 GeV and 100 GeV).
Moving to the right in the figure corresponds to moving forward in time.
Once the ``dots'' are reached, DM annihilation wins over H2 cooling,
and a Dark Star is created.
}
\end{figure}

{\bf Second Criterion: The dark matter annihilation products get stuck inside
the star}.
WIMP annihilation produces energy at a rate per unit volume $Q_{\rm
ann} = \langle \sigma v \rangle \rho_\chi^2/m_\chi \linebreak \simeq  1.2
\times  10^{-29} {\rm erg/cm^3/s} \,\,\, (\langle \sigma v \rangle / (3
\times 10^{-26} {\rm cm^3/s}))  (n/{\rm cm^{-3}})^{1.6} (m_\chi/(100
{\rm GeV}))^{-1}$. In the early stages of Pop III star formation, when
the gas density is low, most of this energy is radiated away 
(\cite{RipamontiMapelliFerrara06}). However,
as the gas collapses and its density increases, a substantial fraction
$f_Q$ of the annihilation energy is deposited into the gas, heating it
up at a rate $f_Q Q_{\rm ann}$ per unit volume.   While neutrinos
escape from the cloud without depositing an appreciable amount of energy,
electrons and photons  can transmit energy to the core.  
We have computed estimates of this fraction $f_Q$ as the core becomes more
dense. Once $n\sim 10^{11} {\rm cm}^{-3}$ (for 100 GeV WIMPs),  e$^-$ and 
photons are trapped and we can take $f_Q \sim 2/3$.

{\bf Third Criterion: Dark matter heating is the dominant heating/cooling mechanism
in the star}.
We find that, for WIMP mass
$m_\chi = 100$GeV (1 GeV), a crucial transition takes place when
the gas density reaches $n> 10^{13} {\rm cm}^{-3}$ ($n>10^9
{\rm cm}^{-3}$).  Above this density, DM heating dominates over
all relevant cooling mechanisms, the most important being
H$_2$ cooling
(\cite{HollenbachMcKee79}).

Figure 2 shows evolutionary tracks of the protostar
in the temperature-density phase plane with DM heating included
(\cite{Yoshida_etal08}),
for two DM particle masses (10 GeV and 100 GeV).  
Moving to the right on this
plot is equivalent to moving forward in time.  Once the
black dots are reached, DM heating dominates over cooling inside the star,
and the Dark Star phase begins.
The protostellar core is 
prevented from cooling and collapsing further.  
The size of the core at this point 
is $\sim 17$ A.U. and its mass is $\sim 0.6 M_\odot$ 
for 100 GeV mass WIMPs.
A new type of object is created, a Dark Star
supported by DM annihilation rather than fusion.


\begin{thebibliography}{99}

\bibitem{SpolyarFreeseGondolo08}
  {D. Spolyar, K.~Freese, \& P.~Gondolo}
  astro-ph/0705.0521, 2008,
\textit{Phys. Rev. Lett.}, 100, 051101


\bibitem{RipamontiAbel05}
  {E.~Ripamonti \& T.~Abel}
  astro-ph/0507130.

\bibitem{BarkanaLoeb01}
  {R.~Barkana \& A.~Loeb} 2001,
\textit{Phys. Rep.}, 349, 125

\bibitem{BrommLarson03}
  {V.~Bromm \& R.~B.~Larson} 2004,
\textit{ARAA}, 42, 79

\bibitem{Yoshida_etal06}
  {N.~Yoshida, K.~Omukai, L.~Hernquist \& T.~Abel} 2006,
\textit{ApJ}, 652, 6

\bibitem{DarkStarsII}
  {D. Spolyar et al}, these proceedings.

\bibitem{NavarroFrenkWhite96}
  {J.~F.~Navarro, C.~S.~Frenk \& S.~D.~M.~White} 1996,
\textit{ApJ}, 462, 563

\bibitem{Blumenthal_etal85}
  {G.~R.~Blumenthal, S.~M.~Faber, R.~Flores \& J.~R.~Primack} 1986,
\textit{ApJ}, 301, 27

\bibitem{AbelBryanNorman02}
  {T.~Abel, G.~L.~Bryan \& M.~L.~Norman} 2002,
\textit{Science}, 295, 93

\bibitem{Gao_etal07}
 {L.~Gao, T.~Abel, C.~S.~Frenk, A.~Jenkins, V.~Springel \& N.~Yoshida} 2007,
\textit{MNRAS}. 378, 449

\bibitem{NatarajanTanO'Shea08}
{A. Natarajan, J. Tan, \& B. O'Shea} 2008,
arXiv:0807.3769 [astro-ph]

\bibitem{FreeseGondoloSellwoodSpolyar08}
{K. Freese, P. Gondolo, J Sellwood \& D. Spolyar} 2008,
arxiv:0805.3540 [astro-ph]

\bibitem{RipamontiMapelliFerrara06}
  {E.~Ripamonti, M.~Mapelli \& A.~Ferrara} 2007,
\textit{MNRAS}, 375, 1399

\bibitem{HollenbachMcKee79}
  {D.~Hollenbach \& C.~F.~McKee} 1979,
\textit{ApJ} Suppl., 41, 555

\bibitem{Yoshida_etal08}
{N. Yoshida, K. Freese, P. Gondolo, \& D. Spolyar} 2008,
in preparation


\end{thebibliography}
\end{document}